\documentclass[a4paper]{revtex4}
\usepackage{graphicx}
\usepackage{fancyhdr}
\usepackage{amsmath}
\usepackage{color}

\newcommand{\perc}{\%}
\newcommand\sss{\scriptscriptstyle}

\begin{document}

\title{Higgs characterisation: NLO and parton-shower effects}

\author{Federico Demartin, Eleni Vryonidou}
\affiliation{ 
 Centre for Cosmology, Particle Physics and Phenomenology (CP3),
 Universit\'e catholique de Louvain,
 B-1348 Louvain-la-Neuve, BELGIUM}

\author{Kentarou Mawatari}
\thanks{speaker}
\email{kentarou.mawatari@vub.ac.be}
\affiliation{Theoretische Natuurkunde and IIHE/ELEM, Vrije Universiteit Brussel,
 and International Solvay Institutes,
 Pleinlaan 2, B-1050 Brussels, BELGIUM}

\author{Marco Zaro}
\affiliation{Sorbonne Universit\'es, UPMC Univ. Paris 06 and CNRS, UMR 7589, LPTHE, 
 F-75005, Paris, FRANCE}

\begin{abstract}
We present the Higgs Characterisation (HC) framework to study the
 properties of the Higgs boson observed at 125~GeV. 
In this report, we focus on CP properties of the top-quark Yukawa
 interaction, and show predictions at next-to-leading order accuracy in
 QCD, including parton-shower effects, for Higgs production in
 association with a single top quark at the LHC. 
\end{abstract}

\maketitle

\thispagestyle{fancy}

\section{Introduction}

The {\it Higgs Characterisation} (HC) presented
in~\cite{Artoisenet:2013puc}, which follows the general strategy
outlined in~\cite{Christensen:2009jx}, provides a framework that allows
one to study the Higgs boson properties, based on an effective field
theory (EFT) approach. 
The EFT lagrangian featuring bosons with various spin-parity assignments
has been implemented in the mass eigenstates in
{\sc FeynRules}~\cite{Alloul:2013bka}, whose output in the
{\sc UFO}~\cite{Degrande:2011ua,deAquino:2011ub} 
can be directly passed to 
{\sc MadGraph5\_aMC@NLO}~\cite{Alwall:2014hca}.
The code is publicly available online in the {\sc FeynRules}
repository~\cite{FR-HC:Online}.  
By employing this framework, we can compute both inclusive cross
sections and differential distributions matched to parton-shower
programs, up to next-to-leading order (NLO) accuracy in QCD, in a fully
automatic way (for the most important spin-0 scenario), and have recently studied all the main Higgs production
channels (gluon fusion, weak vector-boson fusion and
associated production, and $t\bar tH$)~\cite{Maltoni:2013sma,Demartin:2014fia}, as well as the sub-dominant process, associated
production with a single top quark ($tH$)~\cite{Demartin:2015uha}.   
In this report, we focus on $tH$ production, which is particularly
interesting for Higgs characterisation.

\section{Higgs production in association with a single top quark}

\begin{figure}
\center 
\includegraphics[height=0.1\textwidth]{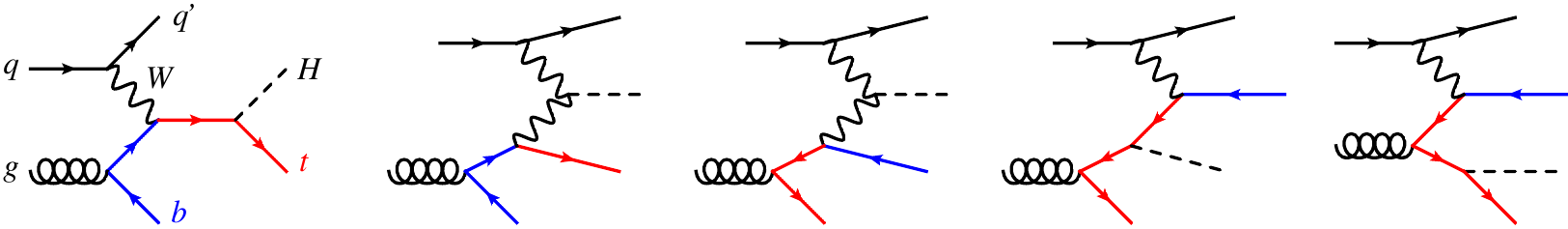}\\[1mm]
\includegraphics[height=0.09\textwidth]{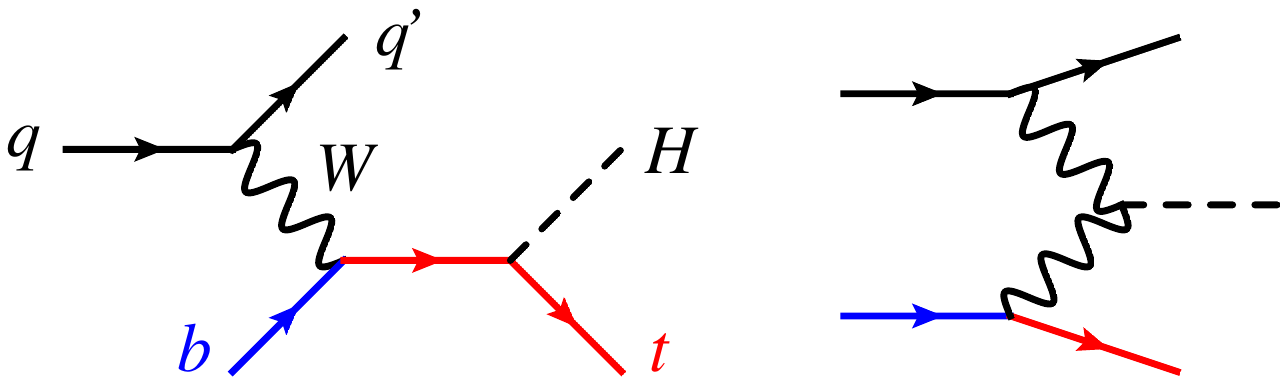}
\caption{LO Feynman diagrams for Higgs production associated with a single top quark via a $t$-channel $W$
 boson 
 in the 4F
 scheme (top) and in the 5F scheme  (bottom).} 
\label{fig:diagram-t}
\end{figure} 

As in single top production in the SM, 
$tH$ production is always mediated by a $tWb$ vertex and therefore it
entails the presence of a bottom quark either in the initial 
($t$-channel and $W$-associated) or in the final state ($s$-channel).

For $t$-channel $tH$ production, diagrams where the
Higgs couples to the top quark interfere destructively with those where
the Higgs couples to the $W$ boson, making cross sections and distributions
extremely sensitive to departures of the Higgs couplings from the SM
predictions.
To assess the possible deviations, reliable predictions and estimates
for the residual uncertainties are
indispensable.  
To this aim, we first provide the SM predictions including QCD corrections
at NLO, paying particular attention to the
uncertainty related to the different flavour schemes,  
so-called 4-flavour (4F) and 5-flavour (5F) schemes; see
fig.~\ref{fig:diagram-t} for the Feynman diagrams. 

\begin{figure}
\center 
 \includegraphics[width=0.45\textwidth]{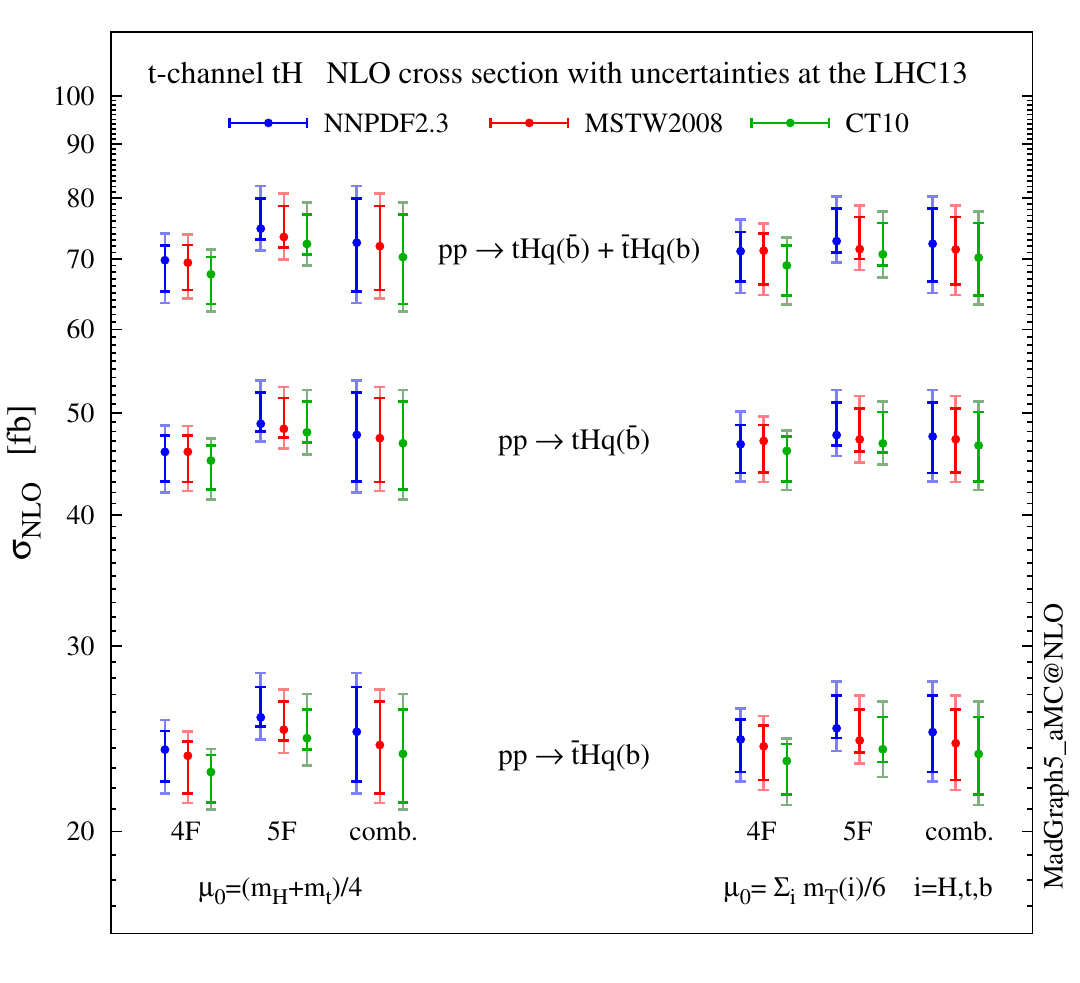}\qquad
 \includegraphics[width=0.42\textwidth]{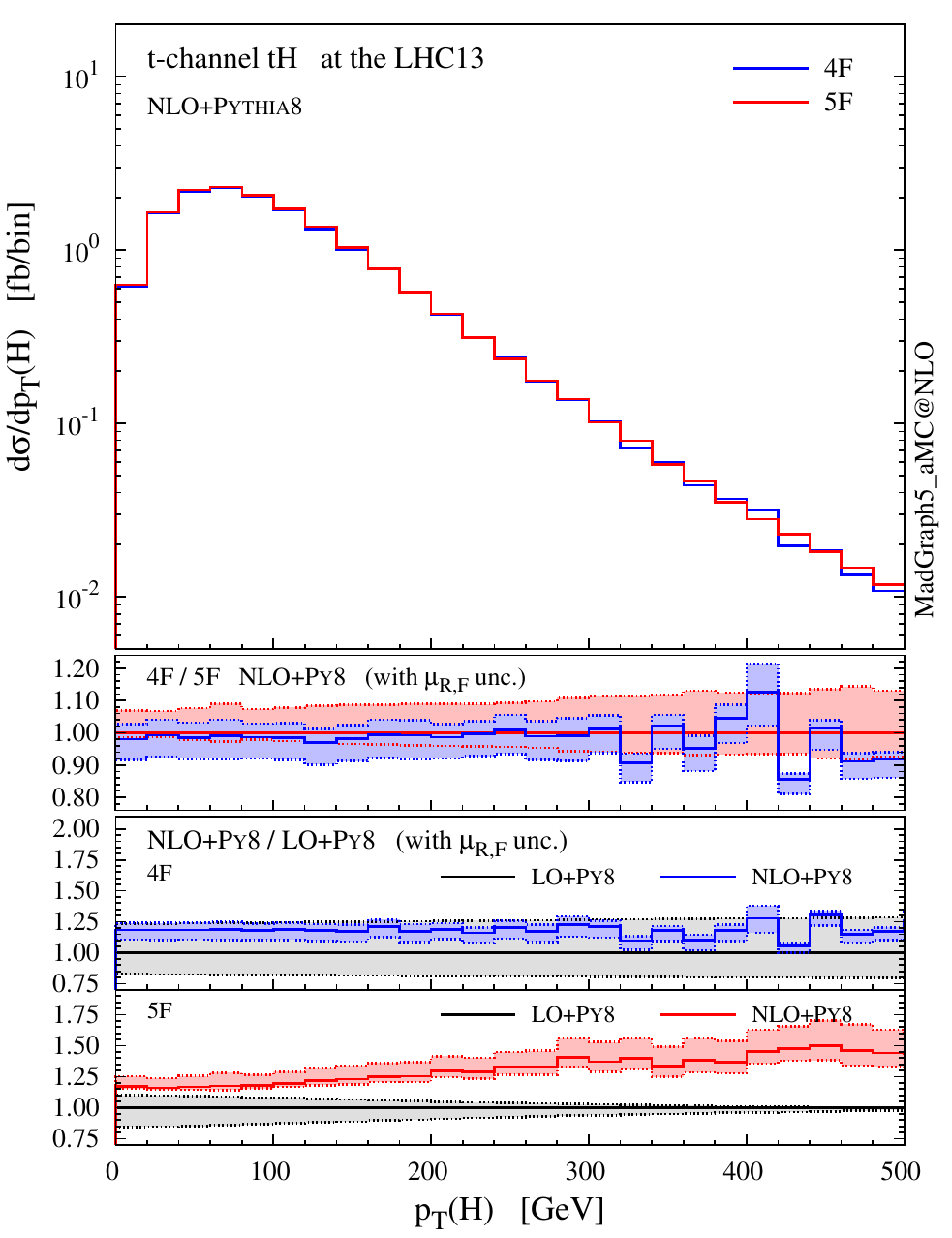}
 \caption{{\it Left:} NLO cross sections with uncertainties for $t$-channel $tH$
 production at the 13-TeV LHC, with different PDF sets.
 For the uncertainties, the inner ticks display the scale (plus combined
 flavour-scheme) dependence $\delta_{\mu(+{\rm FS})}$, while the outer
 ones include $\delta_{{\rm PDF}+\alpha_s+m_b}$.
 The scale dependence is estimated varying the renormalisation and
 factorisation scales by a factor 2 around the static (left) and dynamic
 (right) scale choices.
 {\it Right:} $p_T$ distribution for Higgs boson at NLO+PS accuracy with
 {\sc Pythia8}~\cite{Sjostrand:2007gs}.
The lower panels provide information on the differences between 4F and
 5F schemes as well as the differential $K$ factors in the two schemes.
 See more details in~\cite{Demartin:2015uha}.
 } 
\label{fig:xsec}
\end{figure} 

In {\sc MadGraph5\_aMC@NLO}
the code and events for $t$-channel $tH$ production at hadron colliders,
e.g. in the 4F scheme, can be automatically generated by issuing the following
commands:
\begin{verbatim}
 > import model HC_NLO_X0
 > generate p p > x0 t b~ j $$ w+ w- [QCD]
 > add process p p > x0 t~ b j $$ w+ w- [QCD]
 > output
 > launch
\end{verbatim}
In the HC model, the effective Lagrangian for the Higgs-top quark
interaction reads 
\begin{align}
 {\cal L}_0^t 
   = -\bar\psi_t\big(
         c_{\alpha}\kappa_{\sss Htt}g_{\sss Htt} 
       +i s_{\alpha}\kappa_{\sss Att}g_{\sss Att}\, \gamma_5 \big)
      \psi_t\, X_0 \,,
\label{eq:0ff}
\end{align}
where $X_0$ labels a generic spin-0 particle with CP-violating couplings, 
$c_{\alpha}\equiv\cos\alpha$ and $s_{\alpha}\equiv\sin\alpha$ are related to the 
CP-mixing phase $\alpha$, 
$\kappa_{\sss Htt,Att}$ are real rescaling parameters, and
$g_{\sss Htt}=g_{\sss Att}=m_t/v\,(=y_t/\sqrt{2})$.  
After {\tt launch}, one can modify {\tt param\_card.dat} to change
the parameters, e.g. 
$c_{\alpha}=1 \,,\, \kappa_{\sss Htt}=1 \,$ for the SM case.


\begin{table}
\center
 \caption{NLO cross sections and uncertainties for $t$-channel $tH$
 production at the 13-TeV LHC. 
 {\sc NNPDF2.3} PDFs~\cite{Ball:2012cx} have been used 
 ({\sc NNPDF2.1} for $m_b$ uncertainty in 5F). 
 The integration uncertainty in the last digit(s) (in parentheses) as
 well as the scale (plus combined flavour-scheme) dependence and the
 combined ${\rm PDF}+\alpha_s+m_b$ 
 uncertainty (in $\perc$) are reported.}
\label{tab:xsec_NNPDF23}
\begin{tabular}{rr||llll|llll}
 \hline
    \rule{0pt}{3ex}
    $t$-channel &
  &&$\sigma_{\rm NLO}^{(\mu_0^s)}$~[fb] \quad\quad 
  & $\delta^\perc_{\mu+{\rm (FS)}}$ 
  & $\delta^\perc_{{\rm PDF}+\alpha_s+m_b}$ 
  && $\sigma_{\rm NLO}^{(\mu_0^d)}$~[fb] \quad\quad
  & $\delta^\perc_{\mu+{\rm (FS)}}$ 
  & $\delta^\perc_{{\rm PDF}+\alpha_s+m_b}$ 
    \\[0.7ex]
\hline
    \rule{0pt}{3ex}
  \hspace*{-1em} 4F \enskip \qquad $tH$ &
     && 45.90(7)
     & $^{+3.6}_{-6.3}$
     & $^{+2.3}_{-2.3}$
     && 46.67(8)
     & $^{+4.3}_{-6.1}$
     & $^{+3.2}_{-1.9}$
  \\[0.3ex] 
     \rule{0pt}{3ex} 
 $\bar tH$ &
     && 23.92(3)
     & $^{+4.2}_{-6.6}$
     & $^{+2.5}_{-2.7}$
     && 24.47(5)
     & $^{+4.4}_{-6.8}$
     & $^{+2.5}_{-2.3}$
  \\[0.3ex] 
     \rule{0pt}{3ex} 
 $tH + \bar tH$ & 
     && 69.81(11)
     & $^{+3.2}_{-6.6}$
     & $^{+2.8}_{-2.5}$
     && 71.20(11)
     & $^{+4.3}_{-6.5}$
     & $^{+3.0}_{-2.4}$
  \\[0.7ex] 
\hline 
    \rule{0pt}{3ex} 
  \hspace*{-1em} 5F \enskip \qquad $tH$ &
     && 48.80(5)
     & $^{+7.1}_{-1.7}$
     & $^{+2.8}_{-2.3}$
     && 47.62(5)
     & $^{+7.4}_{-2.2}$
     & $^{+3.0}_{-2.4}$
  \\[0.3ex] 
     \rule{0pt}{3ex} 
 $\bar tH$ &
     && 25.68(3)
     & $^{+6.8}_{-2.0}$
     & $^{+3.4}_{-2.9}$
     && 25.07(3)
     & $^{+7.4}_{-2.1}$
     & $^{+3.2}_{-2.9}$
  \\[0.3ex] 
     \rule{0pt}{3ex} 
 $tH + \bar tH$ & 
     && 74.80(9)
     & $^{+6.8}_{-2.4}$
     & $^{+3.0}_{-2.4}$
     && 72.79(7)
     & $^{+7.4}_{-2.4}$
     & $^{+2.9}_{-2.3}$
 \\[0.7ex] 
 \hline
    \rule{0pt}{3ex} 
  \hspace*{-1em} 4F+5F \enskip \qquad $tH$ &
     && 47.64(7)
     & \scriptsize $ \pm 9.7 $ \normalsize
     & $^{+2.9}_{-2.3}$ 
     && 47.47(6)
     & \scriptsize $ \pm 7.7 $ \normalsize
     & $^{+3.1}_{-1.8}$
  \\[0.3ex] 
     \rule{0pt}{3ex} 
 $\bar tH$ &
     && 24.88(4)
     & \scriptsize $ \pm 10.2 $ \normalsize
     & $^{+3.5}_{-2.6}$ 
     && 24.86(3)
     & \scriptsize $ \pm 8.3 $ \normalsize
     & $^{+3.3}_{-2.3}$
  \\[0.3ex] 
     \rule{0pt}{3ex} 
 $tH + \bar tH$ & 
     && 72.55(10)
     & \scriptsize $ \pm 10.1 $ \normalsize
     & $^{+3.1}_{-2.4}$ 
     && 72.37(10)
     & \scriptsize $ \pm 8.0 $ \normalsize
     & $^{+2.9}_{-2.3}$
  \\[0.7ex] 
 \hline
\end{tabular}
\end{table}


\begin{table}
\center
 \caption{Same as table~\ref{tab:xsec_NNPDF23}, but with MSTW2008
 PDFs~\cite{Martin:2009iq}.} 
\label{tab:xsec_MSTW08}
\begin{tabular}{rr||llll|llll}
 \hline
    \rule{0pt}{3ex}
    $t$-channel &&
  & $\sigma_{\rm NLO}^{(\mu_0^s)}$~[fb] \quad\quad
  & $\delta^\perc_{\mu+{\rm (FS)}}$  
  & $\delta^\perc_{{\rm PDF}+\alpha_s+m_b}$  
  && $\sigma_{\rm NLO}^{(\mu_0^d)}$~[fb] \quad\quad 
  & $\delta^\perc_{\mu+{\rm (FS)}}$  
  & $\delta^\perc_{{\rm PDF}+\alpha_s+m_b}$  
    \\[0.7ex]
\hline
    \rule{0pt}{3ex}
   \hspace*{-1em}  4F \enskip \qquad $tH$ &&
     & 45.91(9)
     & $^{+3.7}_{-6.4}$
     & $^{+2.1}_{-2.0}$
     && 47.00(7)
     & $^{+3.5}_{-6.7}$
     & $^{+1.9}_{-2.1}$
  \\[0.3ex] 
     \rule{0pt}{3ex} 
 $\bar tH$ && 
     & 23.61(3)
     & $^{+3.1}_{-7.9}$
     & $^{+2.4}_{-2.5}$
     && 24.10(5)
     & $^{+4.6}_{-7.1}$
     & $^{+2.2}_{-2.5}$
  \\[0.3ex] 
     \rule{0pt}{3ex} 
 $tH + \bar tH$ && 
     & 69.43(7)
     & $^{+4.0}_{-5.8}$
     & $^{+2.5}_{-1.9}$
     && 71.29(10)
     & $^{+3.8}_{-7.1}$
     & $^{+2.2}_{-2.3}$
  \\[0.7ex] 
\hline 
    \rule{0pt}{3ex} 
  \hspace*{-1em} 5F \enskip \qquad $tH$ &&
     & 48.28(6)
     & $^{+7.0}_{-1.9}$
     & $^{+2.6}_{-2.6}$
     && 47.17(6)
     & $^{+7.0}_{-2.6}$
     & $^{+2.9}_{-2.6}$
  \\[0.3ex] 
     \rule{0pt}{3ex} 
 $\bar tH$  && 
     & 24.99(3)
     & $^{+6.4}_{-2.3}$
     & $^{+2.7}_{-3.1}$
     && 24.41(3)
     & $^{+7.1}_{-2.7}$
     & $^{+3.2}_{-2.8}$
  \\[0.3ex] 
     \rule{0pt}{3ex} 
 $tH + \bar tH$ &&
     & 73.45(8)
     & $^{+7.0}_{-2.3}$
     & $^{+3.0}_{-2.6}$
     && 71.54(7)
     & $^{+7.3}_{-2.1}$
     & $^{+2.8}_{-2.6}$
 \\[0.7ex] 
\hline
    \rule{0pt}{3ex} 
  \hspace*{-1em} 4F+5F \enskip \qquad $tH$ &
     && 47.30(8)
     & \scriptsize $ \pm 9.2 $ \normalsize
     & $^{+2.7}_{-2.0}$
     && 47.18(6)
     & \scriptsize $ \pm 7.0 $ \normalsize
     & $^{+2.9}_{-2.1}$
  \\[0.3ex] 
     \rule{0pt}{3ex} 
 $\bar tH$ &
     && 24.17(4)
     & \scriptsize $ \pm 10.0 $ \normalsize
     & $^{+2.8}_{-2.4}$
     && 24.26(3)
     & \scriptsize $ \pm 7.7 $ \normalsize
     & $^{+3.2}_{-2.5}$
  \\[0.3ex] 
     \rule{0pt}{3ex} 
 $tH + \bar tH$ & 
     && 71.99(11)
     & \scriptsize $ \pm 9.2 $ \normalsize
     & $^{+3.1}_{-1.9}$
     && 71.48(9)
     & \scriptsize $ \pm 7.3 $ \normalsize
     & $^{+2.8}_{-2.3}$
  \\[0.7ex] 
 \hline
\end{tabular}
\end{table}


\begin{table}
\center
 \caption{Same as table~\ref{tab:xsec_NNPDF23}, but with CT10
 PDFs~\cite{Lai:2010vv}.} 
\label{tab:xsec_CT10}
\begin{tabular}{rr||llll|llll}
 \hline
    \rule{0pt}{3ex}
    $t$-channel &&
  & $\sigma_{\rm NLO}^{(\mu_0^s)}$~[fb] \quad\quad 
  & $\delta^\perc_{\mu+{\rm (FS)}}$  
  & $\delta^\perc_{{\rm PDF}+\alpha_s+m_b}$  
  && $\sigma_{\rm NLO}^{(\mu_0^d)}$~{\small [fb]} \quad\quad
  & $\delta^\perc_{\mu+{\rm (FS)}}$  
  & $\delta^\perc_{{\rm PDF}+\alpha_s+m_b}$  
    \\[0.7ex]
\hline
    \rule{0pt}{3ex}
  \hspace*{-1em} 4F \enskip \qquad $tH$ &
     && 45.03(6)
     & $^{+3.4}_{-6.1}$
     & $^{+1.6}_{-2.1}$
     && 46.00(8)
     & $^{+3.3}_{-6.5}$
     & $^{+1.3}_{-1.7}$
  \\[0.3ex] 
     \rule{0pt}{3ex} 
 $\bar tH$ &
     && 22.78(2)
     & $^{+3.8}_{-6.5}$
     & $^{+1.4}_{-1.4}$
     && 23.34(4)
     & $^{+3.8}_{-7.0}$
     & $^{+1.2}_{-2.2}$
  \\[0.3ex] 
     \rule{0pt}{3ex} 
 $tH + \bar tH$ &
     && 67.69(8)
     & $^{+3.9}_{-6.3}$
     & $^{+1.7}_{-1.5}$
     && 69.02(10)
     & $^{+4.5}_{-6.3}$
     & $^{+1.9}_{-1.8}$
  \\[0.7ex] 
\hline 
    \rule{0pt}{3ex} 
  \hspace*{-1em} 5F \enskip \qquad $tH$ &
     && 47.91(6)
     & $^{+7.0}_{-2.2}$
     & $^{+2.7}_{-2.5}$
     && 46.76(6)
     & $^{+7.1}_{-2.0}$
     & $^{+2.5}_{-2.4}$
  \\[0.3ex] 
     \rule{0pt}{3ex} 
 $\bar tH$ &
     && 24.53(2)
     & $^{+6.5}_{-2.5}$
     & $^{+3.7}_{-3.3}$
     && 23.94(3)
     & $^{+7.3}_{-2.7}$
     & $^{+3.7}_{-3.2}$
  \\[0.3ex] 
     \rule{0pt}{3ex} 
 $tH + \bar tH$ &
     && 72.36(9)
     & $^{+6.6}_{-2.4}$
     & $^{+2.9}_{-2.3}$
     && 70.71(8)
     & $^{+7.1}_{-2.5}$
     & $^{+2.7}_{-2.4}$
 \\[0.7ex] 
\hline
    \rule{0pt}{3ex} 
  \hspace*{-1em} 4F+5F \enskip \qquad $tH$ 
     &&& 46.78(6)
     & \scriptsize $ \pm 9.6 $ \normalsize
     & $^{+2.8}_{-2.0}$
     && 46.54(6)
     & \scriptsize $ \pm 7.6 $ \normalsize
     & $^{+2.5}_{-1.7}$
  \\[0.3ex] 
     \rule{0pt}{3ex} 
 $\bar tH$ 
     &&& 23.71(4)
     & \scriptsize $ \pm 10.2 $ \normalsize
     & $^{+3.9}_{-1.3}$
     && 23.70(3)
     & \scriptsize $ \pm 8.4 $ \normalsize
     & $^{+3.8}_{-2.2}$
  \\[0.3ex] 
     \rule{0pt}{3ex} 
 $tH + \bar tH$ 
     &&& 70.29(11)
     & \scriptsize $ \pm 9.8 $ \normalsize
     & $^{+3.0}_{-1.5}$
     && 70.21(9)
     & \scriptsize $ \pm 7.9 $ \normalsize
     & $^{+2.7}_{-1.8}$
  \\[0.7ex] 
 \hline
\end{tabular}
\end{table}

The SM NLO rates and distributions with theoretical uncertainties are
presented in fig.~\ref{fig:xsec} (as well as in tables
~\ref{tab:xsec_NNPDF23}-\ref{tab:xsec_CT10} for the explicit values of
the rates), where the flavour-scheme combined
prediction is defined by 
\begin{align}
 \sigma_{\rm NLO}  = ( \sigma^{+}+\sigma^{-} )/2 \,, \quad
 \delta_{\rm \mu+FS} = ( \sigma^{+}-\sigma^{-} )/2 \,,
\label{eq:enevelope_MU-FS}
\end{align}
with
\begin{align}
 \sigma^{+}  = \max \limits_{\mu \in [\mu_0/2, \, 2\mu_0]} 
 \big\{\, \sigma^{\rm 4F}_{\rm NLO}(\mu) \,,\, 
          \sigma^{\rm 5F}_{\rm NLO}(\mu) \,\big\} \,, \quad
 \sigma^{-}  = \min \limits_{\mu \in [\mu_0/2, \, 2\mu_0]}  
 \big\{ \sigma^{\rm 4F}_{\rm NLO}(\mu) \,,\, 
        \sigma^{\rm 5F}_{\rm NLO}(\mu) \big\} \,.
\end{align}

\begin{figure}
\center
\includegraphics[width=0.45\textwidth]{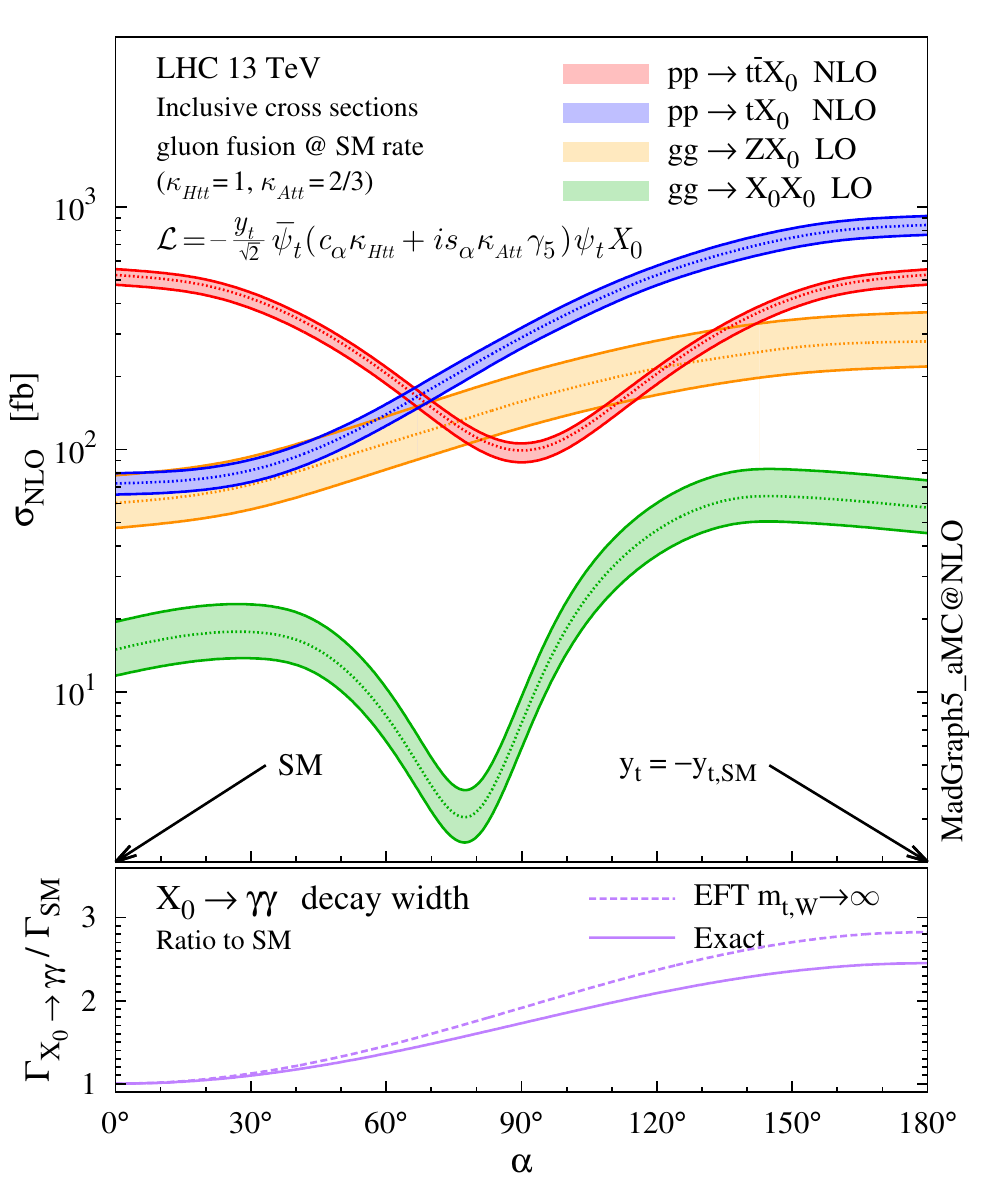}
\caption{NLO (loop-induced LO) cross sections with scale uncertainties for $t\bar tX_0$
 and $t$-channel $tX_0$
 ($ZX_0$ and $X_0X_0$) productions
 at the 13-TeV LHC as a function of
 the CP-mixing angle $\alpha$, where $\kappa_{\sss Htt}$ and 
 $\kappa_{\sss Att}$ are set to reproduce the SM gluon-fusion cross section for
 every value of $\alpha$.
 The ratio of the $X_0\to\gamma\gamma$ partial decay width to the SM
 value is also shown in the lower panel.}
\label{fig:cpmix}
\end{figure}

Finally, we go beyond the SM Higgs coupling to the top quark, and present the dependence on the CP-mixing
angle $\alpha$ for the $tH$ and $t\bar tH$ production cross sections in fig.~\ref{fig:cpmix}.
The nature of the top quark Yukawa coupling also affects the loop-induced 
Higgs coupling to gluons and photons. 
In the figure, to keep the SM gluon-fusion production cross section, 
the rescaling parameters are set to $\kappa_{\sss Htt}=1$ and
$\kappa_{\sss Att}=2/3$.
The LO cross sections for loop-induced $HZ$~\cite{Hespel:2015zea} and 
$HH$~\cite{Hespel:2014sla} production via
gluon fusion are also shown as a reference.


\begin{acknowledgments}
\vspace*{-4mm}
\begin{small}
\noindent
We are grateful to F.~Maltoni for the collaboration on this works.
KM's participation in the workshop was supported in part by the
 Theoretical physics group at the University of Toyama and by the FWO
 travel grant for a long stay abroad.
\end{small}
\end{acknowledgments}


\end{document}